\documentclass[aps,prb,twocolumn,floatfix]{revtex4}
\usepackage{graphicx}
\newcommand{\Av}{\mbox{{\bf A}}}
\newcommand{\Rv}{\mbox{{\bf R}}}

\newcommand{\lv}{\mbox{{\bf l}}}
\begin{document}
\title{Imperfect nanorings with superconducting correlations}
\author{Katarzyna Czajka, Maciej M. Ma\'ska, Marcin Mierzejewski, and {\.Z}aneta {\'S}led{\'z}} 
\affiliation{
Department of Theoretical Physics, Institute of Physics, 
University of Silesia, 40-007 Katowice,
Poland}

\begin{abstract}
The properties of nanoscopic rings with electronic correlations and impurities
are analyzed numerically by means of two complementary methods. Namely, we perform
exact diagonalization of systems up to several lattice sites and 
Bogoliubov--de Gennes--equations studies of systems consisting of a few hundred sites.
We demonstrate how the properties of the systems are affected by various configurations
of impurities for both repulsive and attractive electron--electron interactions.
In the case of attractive interaction we show that the nanoscopic properties 
are mainly determined by the competition between tendencies toward pairing and formation of 
the density waves. Since the impurities act as pinning centers for the density waves, their 
configuration determines the result of this competition.
\end{abstract}
\pacs \ 73.23.Ra,74.78.-w
\maketitle
\section{Introduction}
Recent developments in the fabrication techniques give rise to 
intensive investigations of nanoscopic regime, where the physical 
properties of the system strongly depend on its size. There are 
systems of various geometries, in which the size effects are observed.
In particular, transport properties of nanosystems are of special 
interest. Theoretical analysis of this phenomenon is difficult due to their 
coupling to macroscopic leads. 
Usually, for such systems Coulomb correlations cannot be taken into 
account exactly. From this point of view, some of these systems, 
e.g., nanorings, are very attractive, since their properties can be 
investigated without coupling to the leads. One of the most interesting
features of small metallic rings is the presence of persistent 
currents.\cite{Yang,Buttiker} The currents flow along the rings in equilibrium
state, when a constant external magnetic field is applied. 
Such currents are observed in many experiments, however
their magnitudes are much larger than those predicted theoretically.\cite{duzyprad} 
This discrepancy still remains an 
open problem. It might be attributed to the influence 
of electronic correlations magnetic impurities,\cite{mag_imp} 
or even superconducting fluctuations.
If only one of these 
effects is taken into account, the problem seems to be tractable and 
many theoretical predictions have already been obtained. Generally, it is 
believed that both impurities as well as electronic correlations 
reduce the persistent current. On the other hand, there are some 
indications, that the persistent current can increase, when both these
effects are present. In particular, it has been shown that in the random
potential two interacting electrons  can propagate coherently on a much larger
distance than the one--particle localization length.\cite{shepelyansky}
In some cases, the disorder may lead to $h/2e$  energy levels periodicity,
whereas the corresponding eigenfunctions exhibit a pairing 
effect.\cite{weinmann,aronov} The localization length itself depends on 
the electronic correlations, decreasing (increasing) 
for repulsive (attractive) interactions.\cite{eckern}
For a finite disorder, the persistent currents in the system with repulsive
interactions are larger than those in the system with attractive 
ones.\cite{giamarchi} It is due to the fact that local--density fluctuations
are reduced in the presence of repulsive interactions. 
It is also known, that such counterintuitive
cooperation of correlation 
and disorder can take place also in macroscopic systems of higher 
dimensionality.\cite{byczuk} 
The above argumentation clearly demonstrates that the interplay between disorder and 
electronic correlations is of crucial importance in these systems
and further investigations are needed. 

For many years, the investigated rings have been made out of normal metals. 
Only recently, the technological progress allowed
investigation of superconducting nanowires.\cite{tinkham} 
As the sizes 
of such systems can be comparable to the coherence length, a question 
concerning the possible onset of superconductivity became very interesting.
The experiments show that, in sufficiently thin nanowires, the
superconductivity does not occur.\cite{markovic} The suppression of superconductivity
is usually attributed to the destruction of the phase coherence by quantum
phase slips.\cite{slipy1,slipy2}
The spatial confinement originating from the geometry of a nanowire 
is responsible for an inhomogeneity of the superconducting order 
parameter.\cite{han}
The physical properties of mesoscopic superconducting rings are presently 
intensively investigated. For extremely type II superconductors 
fabrication of nanorings should be possible and will probably be the
subject of future experiments. This problem contains interesting
physics, because both superconducting and one--electron 
persistent currents may occur in such systems.
Moreover, one may expect that phenomena typical for low 
dimensional correlated systems, e.g., charge density waves (CDW),
may be present as well. 
On the other hand, magnetic flux
strongly affects the CDW ground state in ring--shaped systems\cite{nathanson} and 
may even lead to its destruction.\cite{yi} The CDW order could be strongly affected 
also by the impurities, as they play the role of pinning centers.\cite{lee}
 
The aim of this paper is a detailed investigation of the nanoscopic rings with pairing 
correlations and impurities. In particular, we focus on the influence of
impurities on competition between superconductivity and the CDW. 
In the first part of this contribution, we consider rings small enough to be investigated 
within the exact diagonalization methods, leading to rigorous results.
Then, we compare these results with the ones obtained for much larger systems with the 
help of the Bogoliubov--de~Gennes equations. 

The outline of this paper is as follows. In Sec. II we recall main results concerning the influence
of the Coulomb interactions on the persistent current in nanorings. 
In Sections III and IV we investigate nanorings with pairing interactions in the presence of impurities.
In Sec. III we present rigorous results obtained from an exact diagonalization study, whereas Sec. IV
contains similar results obtained with the help of BdG equations for rings containing up to a few hundred sites.
Finally, in Sec. V we summarize our results.

\section{Correlations and impurities in nanorings}

We start our investigations with small, up to 12--sites, rings described by the 
Hubbard Hamiltonian,
\begin{equation}
H_{\rm Hubb}=-t\sum_{\langle i,j\rangle,\sigma} e^{i \theta_{ij}} a^\dagger_{i\sigma} a_{j\sigma}
+U\sum_i n_{i\uparrow}n_{i\downarrow},
\label{hubbar}
\end{equation}
where $a^\dagger_{i\sigma}\ (a_{i\sigma})$ creates (annihilates) an electron on site $i$
with spin $\sigma$, $U$ is the on--site electron--electron interaction, and 
$n_{i\sigma}=a^\dagger_{i\sigma} a_{i\sigma}$. $t$ is the nearest--neighbor hoping integral 
in the absence of magnetic field ($t>0$) and $e^{i \theta_{ij}}$ is the Peierls phase
factor that describes the orbital response of the system to an external magnetic field:
\begin{equation}
\theta_{ij}=
\frac{2 \pi}{\Phi_0} \int^{\Rv_{i}}_{\Rv_{j}}
\Av\cdot d\lv, 
\label{peierls}
\end{equation}
where $\Phi_0=hc/e$ is the flux quantum.
This Hamiltonian has exactly been diagonalized with the help of 
the Lancz\"os algorithm. It is one of the most effective computational
tools for searching the ground state and some low laying excited
states of a finite system. From the ground state, we can compute
all static and dynamic properties, and in this sense, we obtain a
complete characterization of a model at low temperatures. 
At zero temperature the flux--induced current $I$ is calculated as
\begin{equation}
I=-\frac{dE_0}{d\Phi},
\label{current}
\end{equation}
where $E_0$ is the ground state energy and $\Phi$ is the magnetic flux
piercing the ring. At finite temperature in Eq. (\ref{current}) one should use 
the free energy ${\cal F}$ instead of the ground state energy. Unfortunately, 
the Lancz\"os method gives only a few lowest eigenenergies and, therefore,
the calculations are restricted to relatively low temperatures. On the 
other hand, for smaller systems other methods enabled us to find all the eigenenergies of 
the Hamiltonian, and the resulting current can be obtained for an arbitrary
temperature. 
\begin{figure}
\includegraphics[width=8cm]{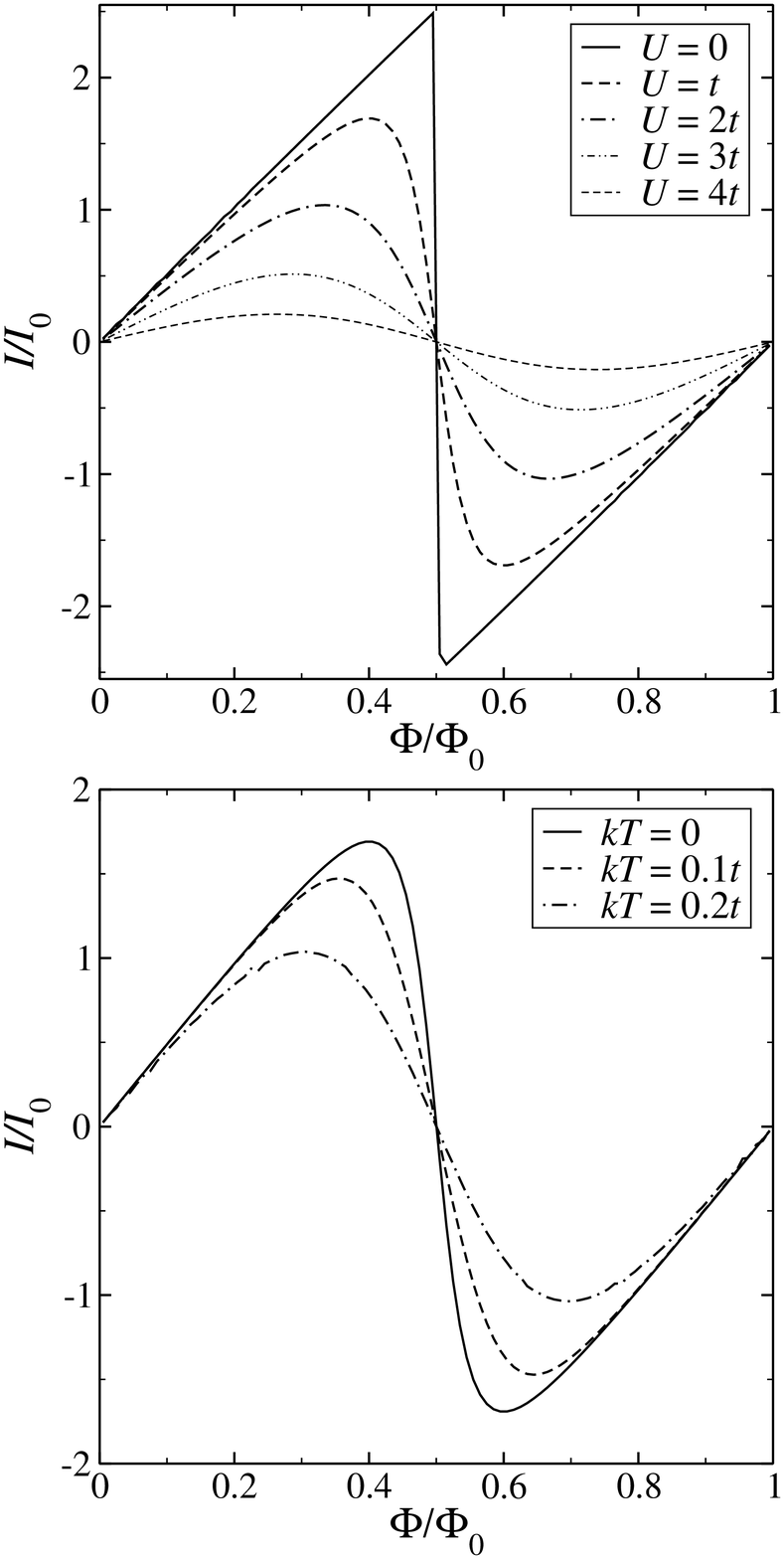}
\caption{
Flux dependence of the persistent current for a ring containing 10 lattice sites 
with repulsive interaction and in the
absence of impurities. Results presented in the upper panel
have been obtained for $T=0$ and various values of $U$, as indicated in the legend.
The lower panel shows  results obtained for $U=t$ at various temperatures.
We have denoted $I_0=t/\Phi_0.$
}
\label{fig1}
\end{figure}
In Fig. 1 we demonstrate how the persistent currents are destroyed by
the Coulomb repulsion (upper panel) and by the increase of the temperature (lower panel).
It has also been shown that the  persistent currents are reduced in the presence of the
thermal equilibrium noise.\cite{dajka} 
These results are intuitive and well known and, therefore, we will not discuss
them here. They are presented only for comparison with the results discussed further.

In order to account for the presence of nonmagnetic impurities we extend the Hubbard 
Hamiltonian
\begin{equation}
H=H_{\rm Hubb}+\sum_i w_i \left( n_{i\uparrow}+n_{i\downarrow}\right),
\label{imp}
\end{equation}
where $w_i$ is the potential of an impurity at site $i$.

\begin{figure}
\includegraphics[width=8.5cm]{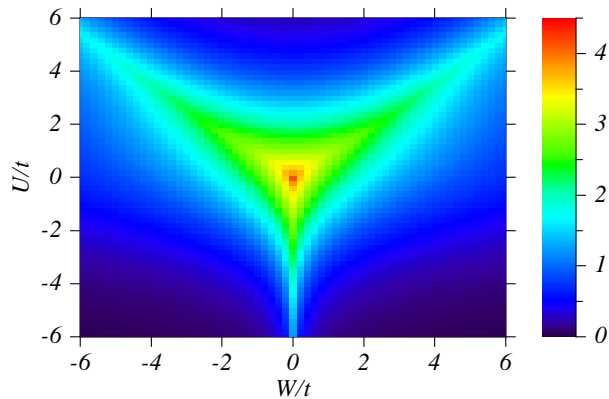}
\caption{
Dependence of the $I_{\rm max}/I_0$ on the electron--electron interaction ($U$) and
the impurity potential ($W$) for a 6--sites ring with a single impurity.
}
\label{fig2}
\end{figure}
\begin{figure}
\includegraphics[width=8cm]{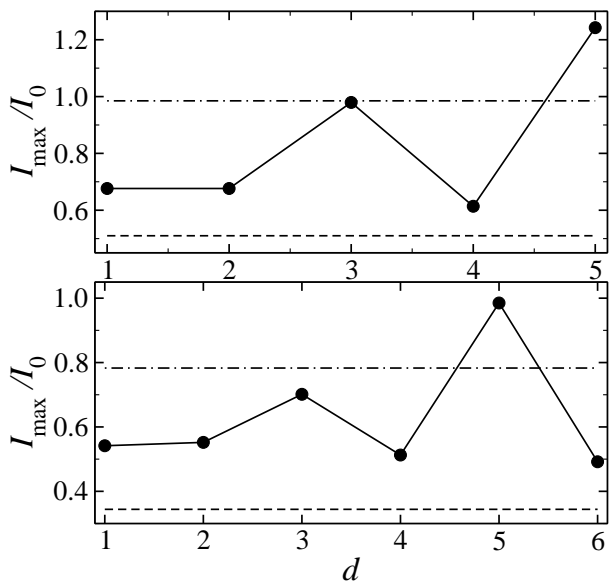}
\caption{
$I_{\rm max}$ calculated for a ring with two impurities as a function of the distance between them.
We have taken $U=3t$, $W=-4t$ and $T=0$. 
The lower (upper) panel corresponds to the 12--sites (10--sites) ring. 
In both the panels horizontal lines indicate $I_{\rm max}$ in the absence of impurity (the lower one)
and in the presence of a single impurity (the upper one). 
}
\label{fig3}
\end{figure}

We start our investigations with a single impurity, i.e., $w_i=\delta_{i0}W$, in a small ring. 
The first question that arises in this case concerns the impact of the electron correlations and impurity
on the magnitude of the persistent current. To answer this question, for a wide range the potentials $U$ and $W$ we have
found a magnetic flux that produces the maximal value of the persistent current ($I_{\rm max}$). 
Fig. \ref{fig2} shows how $I_{\rm max}$ depends on $U$ and $W$ for the case of half--filling, i.e.,
when the number of electrons is equal to the number of sites. One can see that there is a 
$W \rightarrow -W $ symmetry. It is an obvious result of the particle--hole symmetry of the Hubbard model.
In the case of the attractive electron--electron interaction the maximal value of $I_{\rm max}$
takes place for $W=0$, i.e., in the absence of  impurities. Contrary to this result,
for repulsive on--site interaction ($U>0$), $I_{\rm max}$ takes on the maximal value in the presence
of impurity, when $|W|$ is slightly larger than $U$. It means that for a fixed value of the 
impurity potential the maximum of $I_{\rm max}$ corresponds to the finite repulsive interaction and
this result holds independently of the sign of the impurity potential.  Similarly,
in Ref. \onlinecite{giamarchi} it has been shown that in a disordered ring, for the repulsive interaction
the persistent current is larger than for the  attractive one. It originates from the fact that repulsive
interaction reduces impurity--induced density fluctuations, whereas attractive interaction may lead to 
CDW with impurities acting as pinning centers.

One meets much more interesting situation in a case of many impurities.
In particular, a question arises whether an impurity added to the previously considered ring, 
leads to a further enhancement or a reduction of the persistent current. 
The answer to this question strongly depends on the relative positions of the impurities. 
Fig. \ref{fig3} shows $I_{\rm max}$ as a function of the
distance $d$ between two impurities, i.e., for $w_i=W(\delta_{i0}+\delta_{id})$. 
One can see, that for two impurities located at the
nearest and the next nearest neighbors the persistent current 
is smaller than in the case of a single impurity but larger than for $W=0$. 
Then, one can see an oscillatory character of this dependence with increasing amplitude. 
Namely,  $I_{\rm max}$ is enhanced (reduced) when the distance is odd (even) multiple of the
lattice constant. For sufficiently large distance between impurities the persistent current may
even exceed $I_{\rm max}$ obtained for the case of a single impurity. The obtained oscillatory behavior
may be attributed to the density oscillations induced by impurities (Friedel oscillations).
It is expected\cite{eckern} that these oscillations asymptotically decay with the distance $x$ as 
$\cos(2 k_F x+\eta) x^{-\delta}$. Therefore, oscillations originating from different impurities 
may interfere. For the half--filled case, $2 k_F=\pi$ and the oscillations produced by impurities 
separated by odd (even) number of the lattice constants interfere destructively (constructively).

So far, we have focused on the case of a repulsive electron--electron interaction, and we have shown
that the configuration of impurities is of vital importance for the magnitude of the persistent 
current. In the case of the attractive interaction one may expect that this effect should be even more
pronounced, since in such a model  a CDW instability occurs also in the absence of impurities.
Then, impurities may enhance this ordering acting as pinning centers. This problem will be investigated in the 
following section. 

\section{Nanorings with pairing interaction}

\begin{figure}
\includegraphics[width=8cm]{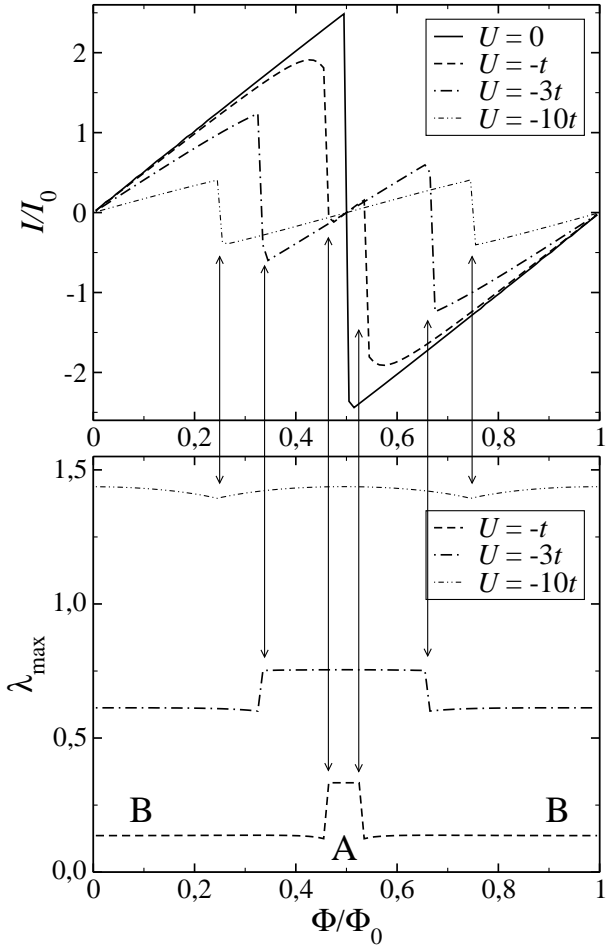}
\caption{The upper panel shows the flux dependence of the persistent
current for 10--sites ring in the absence of impurities. The curves correspond to
various values of the pairing potential. The lower panel shows the pair susceptibility
for the same parameters as in the upper panel. The  arrows indicate on the 
coincidence between the abrupt change of pair susceptibility and the reversed circulation of the
persistent currents.
Letters A and B mark regimes of large and small values of $\lambda_{\rm max}$, respectively. 
}
\label{fig4}
\end{figure}

We start with the attractive Hubbard model without impurities. The upper panel in Fig. \ref{fig4}
shows the persistent current as a function of the magnetic flux for different
values of the on--site pairing potential $U$. As the interaction increases, the system
evolves towards a state where the persistent current exhibits 
$\Phi_0/2$  periodicity. 
Simultaneously, one may observe a reduction of the magnitude of the persistent  
currents when the pairing increases. 
The change of periodicity 
may be a signature of a current made out of carriers having charge 
$2e$,\cite{byers,yang}
although one does not expect occurrence of a superconducting phase in such a small system. 
Similar results have recently been obtained for boson--fermion model.\cite{ranninger}
However, the change of periodicity of the persistent current is not necessarily related
to the pairing interaction. In particular, 
for a genuinely strong on--site repulsion, the system consisting of
$N_e$ electrons shows $\Phi_0/N_e$ and $\Phi_0/2$ periodicities.\cite{kusmartsev,koskinen} 
Therefore, it would be important to distinguish between the possible mechanisms, that may be 
responsible for the change of periodicity. In order to perform this task, we calculate
the pairing correlation function for local Cooper pairs. Usually, one calculates 
the susceptibility of the form\cite{dagotto}
\begin{equation}
\chi_{\rm sup}=\frac{1}{N} \sum_{ij}
\left( \langle \hat{\Delta}_i \hat{\Delta}^{\dagger}_j \rangle - 
\langle a_{i \uparrow} a^{\dagger}_{j \uparrow} \rangle 
\langle a_{i \downarrow}  a^{\dagger}_{j \downarrow} \rangle
\right),
\end{equation}
where the Cooper pair creation operator is given by $\hat{\Delta}^{\dagger}_i=a^{\dagger}_{i \downarrow}
a^{\dagger}_{i \uparrow} $. The increase of this quantity indicates that pairing correlations 
are enhanced. However, in the presence of magnetic field, we cannot directly use this form of the
susceptibility. Instead, we need a gauge--invariant quantity. This will ensure that the susceptibility
will show the same periodicity as the system under investigation. Therefore, we construct a hermitian 
matrix  
\begin{equation}
\chi_{ij}= \langle \Delta_i \Delta^{\dagger}_j \rangle - 
\langle a_{i \uparrow} a^{\dagger}_{j \uparrow} \rangle 
\langle a_{i \downarrow}  a^{\dagger}_{j \downarrow} \rangle,
\end{equation}
and investigate its eigenvalues. They are gauge--invariant and posses the same periodicity
as the energy spectrum.
In an infinite system, the superconducting 
instability corresponds to the divergence of $\chi_{\rm sup}$. In such a case the maximal eigenvalue
of $\chi_{ij}$, $\lambda_{\rm max}$,  diverges as well. Therefore, in the presence of magnetic field we use 
$\lambda_{\rm max}$ as a quantity that probes the tendency towards the formation of the paired state.
The lower panel of Fig. \ref{fig4} shows $\lambda_{\rm max}$ as a function of magnetic field. 
One can see that $\lambda_{\rm max}$ strongly depends on the magnetic flux. This quantity
is maximal for exactly the same values of the flux 
(regime that is marked by ``A'' in Fig. \ref{fig4}), for which the persistent currents are
modified by the pairing correlations. Therefore, we identify these regimes as 
precursors to the superconducting state. For a weak attraction the enhancement of 
 $\lambda_{\rm max}$ in these regimes is pronounced. As the interaction strength increases, these
regimes become wider, however, simultaneously the field--dependence of $\lambda_{\rm max}$ gradually vanishes.
It means that for a weak interaction superconducting correlations are enhanced by specific values
of the magnetic flux, whereas for strong coupling the pairing tendency is independent of the
flux. This is a remnant of the Little and Parks results obtained for macroscopic thin 
superconducting films.\cite{little,gennes} 
This problem will be discussed in more detail in the following section within the Bogoliubov--de Gennes
approach. Additionally, as one may expect, the maximal value of $\lambda_{\rm max}$ increases when 
the paring interaction becomes stronger, supporting our interpretation of this quantity.

\begin{figure}
\includegraphics[width=8cm]{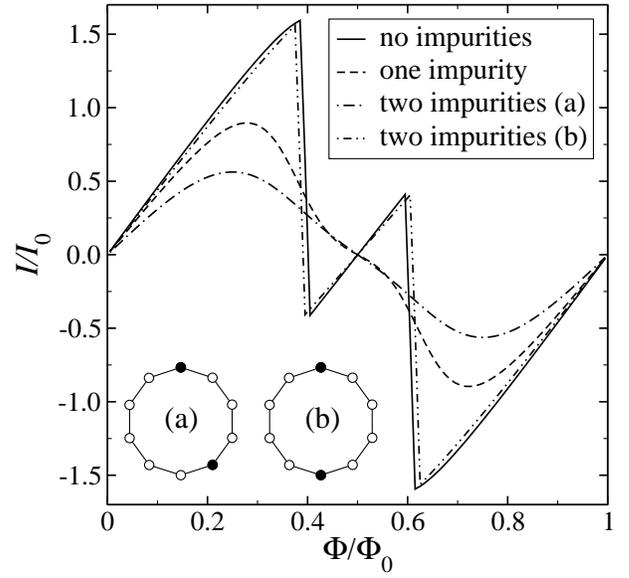}
\caption{
The flux dependence of the persistent current for 10--sites
ring with attractive interaction $U=-2t$ and impurity potential
$W=0.5t$. The curves have been obtained for a system without 
impurities, with a single impurity, and two impurities in configurations
presented in the insets.  
}
\label{fig5}
\end{figure}

Now, we extend the analysis taking into account the impurities. It has already been shown
(see Fig. \ref{fig2}) that here, in contradistinction to the case of the repulsive interaction, 
a single impurity always reduces the persistent current. Playing a role of a CDW pinning center,
it stabilizes density waves, which compete with pairing. It shows up as a vanishing of the 
regime of reverse circulation of the persistent current. However, in the presence of
many impurities they can reduce as well as enhance the persistent current, depending on 
their configuration. Fig. \ref{fig5} presents the flux dependence of the persistent currents
for some configurations of two impurities.  

One can see that a single impurity always reduces the persistent current and destroys
the tendency towards formation of the paired state. 
On the other hand, when an additional impurity is introduced into the system, the 
persistent current can be significantly larger then
in the case of a single impurity. This, however, depends  on the relative position of the impurities.
Similarly to the case of repulsive interaction, when the distance between the impurities is an odd (even) 
multiple of the lattice constants, the persistent current and the pairing tendency are enhanced 
(reduced). This is a result of the competition between the CDW order and the formation of Cooper pairs.
For the half--filled case the electron density in the CDW state oscillates with
the wave--vector $\pi$. Therefore, depending on configuration of the impurities, the density--waves pinned 
by them can interfere constructively or destructively,  increasing  or reducing the CDW order. 
In the first case, the persistent current is less than in the presence of a single impurity and 
the tendency towards formation of the paired state is almost destroyed. In the latter case
the persistent current can be as large as in the clean system. One can see from Fig. \ref{fig5} that
for some configurations of impurities the persistent current can be almost indistinguishable from
that obtained for the clean system.

\section{Nanorings of finite width}
\subsection{The formalism}

In the proceeding sections we have analyzed one--dimensional systems
consisting of several sites only. This limitation originated from the Lancz\"os 
method, that we have used to diagonalize the Hamiltonian. In this section
we extend this analysis and account for finite--width rings consisting of a few 
hundred lattice sites. In such a case we cannot use the exact diagonalization method
and, therefore, the interaction term is analyzed at the mean--field level. In particular,
we  decouple this  term in the following way: 
\begin{eqnarray}
U\sum_i n_{i \uparrow} n_{i \downarrow} &\simeq&
U\sum_i \left(\langle n_{i \uparrow} \rangle n_{i \downarrow} +
 n_{i \uparrow} \langle n_{i \downarrow} \rangle \right)  \nonumber \\
&& + U \sum_i \Delta_i a^{\dagger}_{i \uparrow} a^{\dagger}_{i \downarrow} + {\mathrm h.c.},
\label{mf}
\end{eqnarray}
where the superconducting order parameter reads 
$\Delta_i= \langle a_{i \downarrow} a_{i \uparrow}\rangle $.
The first term on the rhs of Eq. \ref{mf} is responsible  for the formation of density waves. For the 
negative $U$, the second term leads to isotropic $s$--wave superconductivity.
In the following, we assume that there is no magnetic ordering, i.e., 
$\langle n_{i \uparrow}\rangle=\langle n_{i \downarrow}\rangle=\bar{n}_i$.
As the system under investigation is inhomogeneous, both the superconducting and CDW order 
parameters are site--dependent
and have to be determined in a self--consistent way from the Bogoliubov--de Gennes (BdG) 
equations.\cite{gennes}  This approach has most commonly been used for the investigation of the
vortex structure in macroscopic superconducting  systems.\cite{wang,scaf,my}
We introduce a set of new fermionic operators
$\gamma^{\left( \dagger \right)}_{n \sigma}$:
\begin{eqnarray}
a_{i \uparrow}&=& \sum_{l} u_{il} \gamma_{l \uparrow}
-v^*_{il}\gamma^{\dagger}_{l \downarrow}, \nonumber \\
a_{i \downarrow}&=& \sum_{l} u_{il} \gamma_{l \downarrow}
+v^*_{il}\gamma^{\dagger}_{l \uparrow}, \nonumber
\end{eqnarray}
where
\begin{equation}
\sum_{j}\left(
\begin{array}{cc}
{\cal H}_{ij} & U \Delta_{i} \delta_{ij} \\
U \Delta_{i}^* \delta_{ij} & -{\cal H}_{ij}^* 
\end{array}
\right)
\left(
\begin{array}{c}
u_{jl} \\
v_{jl}
\end{array}
\right)
=E_l
\left(
\begin{array}{c}
u_{il} \\
v_{il}
\end{array}
\right).
\label{BdG}
\end{equation}
Here, the single particle Hamiltonian is given by
\begin{equation}
{\cal H}_{ij}=-t\delta_{i+{\bf \delta},j}e^{i \theta_{ij}}+
\left(U \bar{n}_i+w_i-\mu\right)\delta_{ij},
\end{equation}
where $\mu$ is the chemical potential.
The superconducting oder parameter is determined self--consistently by:
\begin{equation} 
\Delta_{i}=-\sum_l
u_{il}v_{il}^*
\tanh\left(\frac{E_l}{2kT}\right).
\label{BDGSC}
\end{equation} 
Also, the local electron concentration is calculated self--consistently in the following way:
\begin{equation}
\bar{n}_i=\sum_{l}
|u_{il}|^2 f(E_l) +|v_{il}|^2 f(-E_l),
\label{ni}
\end{equation}   
where $f$ is the Fermi distribution function.
This quantity allows one to define the CDW order parameter:
\begin{equation}
\Omega_i=(-1)^i\:\left(\bar{n}_i-\bar{n}\right),
\end{equation}
where $\bar{n}=1/N \sum_j \bar{n}_j $ is the average concentration of electrons in the
ring.
Up to this point, we have investigated total current flowing along
the one--dimensional ring. Now, we investigate the current distribution
in a ring of a finite width.     
We follow the procedure described in Ref. \onlinecite{current}.
Namely, the current from site $i$ to the neighboring site $j$ reads:
$I_{ij}=\langle \partial H / \partial \bar{A}_{ij} \rangle$, where  $\bar{A}_{ij}$
is the integral of the vector potential between sites $i$ and $j$. Then,
it is easy to show that
\begin{eqnarray}
I_{ij} &=&- \frac{2et}{\hbar c}\: {\mathrm Im} \left[e^{i \theta_{ij} }
\sum_{l}  \tanh\left(\frac{E_l}{2kT}\right)  \right. \nonumber  \\
&& \times \left. \left( v_{il}v^*_{jl} -u^*_{il}u_{jl} \right)
 \right].
\end{eqnarray}

\subsection{Numerical results for a clean system.}

\begin{figure}
\includegraphics[width=8cm]{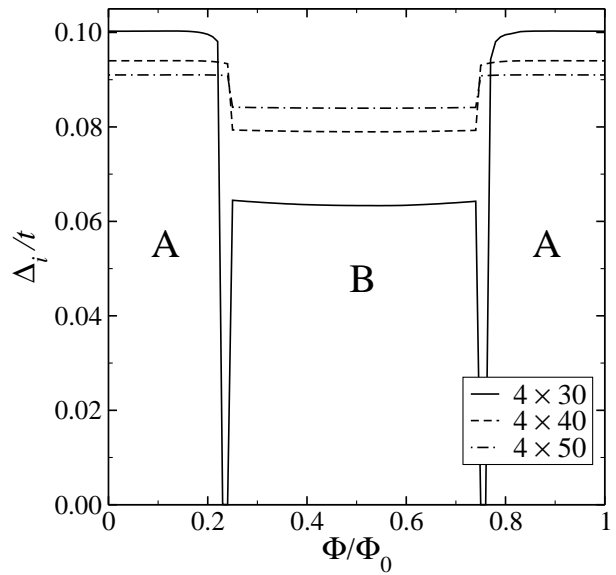}
\caption{The flux dependence of the superconducting order parameter for
nanorings of various size without impurities. $U=-1.25t$ and $kT=0.01t$ have been used.
We have marked regions A and B (see the text).}
\label{fig6}
  \end{figure}

\begin{figure}
\includegraphics[width=8cm]{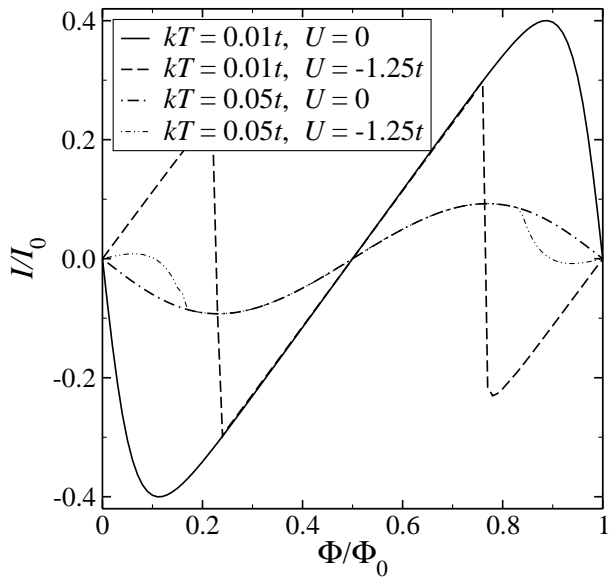}
\caption{Comparison of the persistent current in the presence and in the absence of the pairing
potential for $4\times 30$ ring without impurities.}
\label{fig7}
\end{figure}

We have solved the BdG equations for rings of sizes $4\times M$, where $M=30,40,50$.
In order to determine how the properties of the ring depend on its size, we have
started our investigations with a system without impurities. Due to the small size of
the system the influence of its edge is non--negligible and results in 
an inhomogeneity of $\Delta_i$.\cite{inhomo} However, the differences of flux dependence
of $\Delta_i$ between various lattice sites are of quantitative character only.
Therefore, we present results for one particular site, that is close to the midway 
point between the ring's edges. In Fig. \ref{fig6} we show the magnitude of the superconducting 
order parameter as a function of the applied magnetic flux 
at low temperature and  for various $M$.
One can see that the flux dependence of the order parameter increases with 
decreasing circumference of the ring.  
At low temperature, the magnitude of $\Delta_i$
takes on two different values. In analogy to Fig. \ref{fig4} we denote the regimes of high and low 
values of $|\Delta_i|$ as A and B, respectively. Comparing Figs.  \ref{fig6} and \ref{fig7},
one can note opposite directions of the supercurrent's circulation 
respectively to the normal--state persistent current in regimes A and B. Moreover,
the regime A (B) becomes wider (narrower) when the circumference increases.
In the case of infinite circumference of the ring, the normal--state persistent 
currents vanish, regimes A and B become indistinguishable, and the system exactly 
exhibits $\Phi_0/2$ periodicity. 
The flux dependence of the superconducting order parameter is similar to that of
the pair susceptibility calculated for smaller rings with the help of the Lancz\"os method. 
However, in the latter case 
the regime A is much narrower due to much smaller ring's size. It is well known that 
the mean--field approximation is inappropriate for low--dimensional systems. 
However, it seems that this simple approach correctly describes the persistent currents in
small rings with weak local attraction. 

In Fig. \ref{fig7} we compare persistent currents in the normal and 
superconducting states. It is interesting that in regime B, the persistent current
is the same in the presence and in the absence of the pairing interaction. Again,
similar behavior has been obtained in the exact diagonalization study, presented in
the proceeding section. In the upper panel of Fig. \ref{fig4}  one can see that for
weak attraction the persistent current in the regime B hardly depends on $U$.     

\begin{figure}
\includegraphics[width=8cm]{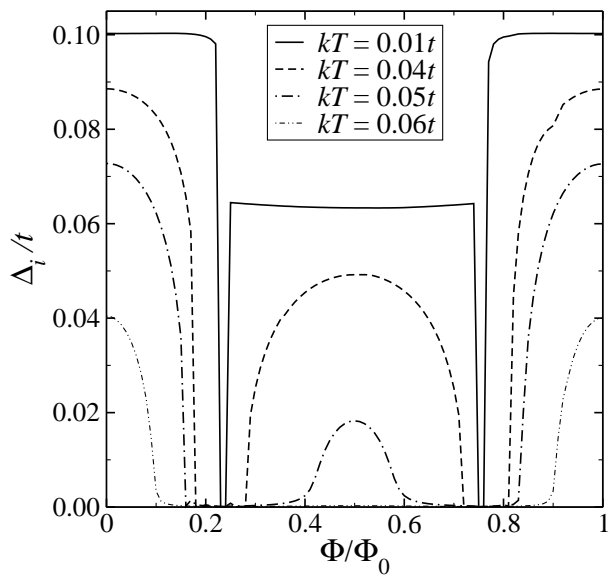}
\caption{The same as in Fig. \ref{fig6}, but at various temperatures.}
\label{fig8}
\end{figure}

Finally, we investigate how the flux dependence of the order parameter depends on temperature
in the case of nanoscopic rings. It has been well known since the famous experiment of
Little and Park that  properties of a superconducting thin film deposited on an 
insulating cylinder depend on an axial magnetic field.\cite{little} In particular,
the transition temperature is a periodic function of the magnetic flux with a period
$\Phi_0/2$.\cite{gennes} This experiment has been carried out for a macroscopic system. Fig.~\ref{fig8}
shows similar dependence for a small ring, where the finite--size effects are important.
As one may expect also in this  case the transition temperature is flux dependent. However,
there is a visible deviation from the $\Phi_0/2$ periodicity. Generally, in the regime B the
superconducting order parameter is less than in the regime A. When the temperature increases, 
superconductivity  first disappears in the regime B and then in the regime A.
The same effect can be observed when the temperature is fixed but the pairing potential is 
reduced.\cite{zhu}
However, in contradistinction to the macroscopic film, in the present case, the vanishing of superconductivity
does not correspond to the vanishing of current. Comparing Figs. \ref{fig7} and \ref{fig8} one can note
that the current remains finite also for  $\Delta_i=0$. 

\subsection{Impurities}

In the case of small one--dimensional rings the presence of impurities significantly changes
the persistent current. In the following, we show that the effect is very important
in much larger rings as well. This holds also if the concentration of impurities
is relatively low. In particular, similarly to the one-dimensional case, the presence
of a single impurity strongly reduces superconductivity. It originates from the
pinning of the density wave, that competes with the superconductivity. In the case
of many impurities, both the CDW and superconducting orders may coexist in the ring.
However, the competition between CDW and superconductivity leads to a spatial separation
of regions, where these orders dominate. The distribution of these regions is determined 
by the configuration of impurities. In the vicinity of impurities, the CDW order dominates.
This enhancement of the CDW order is strongest, if the impurities are located in such 
a way, that the pinned density waves are in phase. Otherwise, the effect of impurities is
much less important. This is a result of vanishing of the CDW order parameter somewhere in
between the impurities. In the region of vanishing CDW order, the superconductivity is
strongly enhanced. This effect is similar to that obtained for the vortex structure 
in Ref. \onlinecite{my}, where the $d$--wave superconductivity competes with $d$--density waves.
	
\begin{figure}
\includegraphics[width=9.8cm]{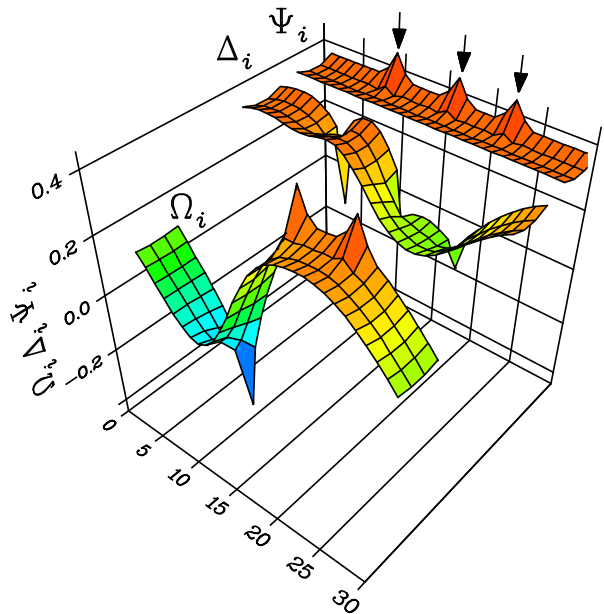}
\caption{The spatial distribution of the CDW (the firs stripe) and superconducting
(the second stripe) order parameters, $\Omega_i$ and $\Delta_i$, respectively, 
in $4\times 30$ ring for $U=-4t$, $W=-2t$ at $kT=0.01t$. 
The third stripe shows $\Psi_i=\sqrt{\Delta_i^2+\Omega_i^2}$. Arrows above the third 
stripe indicate the sites, where impurities are located. For the sake of legibility, 
the ring has been cut and unfolded.}
\label{fig9}
\end{figure}

Fig. \ref{fig9} shows the spatial distribution of the CDW and superconducting
order parameters. The positions of the impurities are indicated by vertical arrows.
In order to prove that the competition between these orders is responsible for
their spatial distribution, we present also the sum of squares of the order parameters,
$\Psi_i=\sqrt{\Delta_i^2+\Omega_i^2}$. This quantity is almost constant over the whole
ring (except for a very close vicinity of the impurities), what confirms our interpretation.

\section{Conclusions}
We have presented numerical analysis of a nanoscopic ring pierced
by an external magnetic flux. A large number of factors affecting
the properties of the ring has been taken into account. In particular, 
we have rigorously treated the electron--electron interaction, demonstrating
its role in the reduction of the persistent current. This effect, however,
can be much less effective when impurities are introduced into the ring. It has been shown
that the restoration of the persistent current occurs as a result of a reduction
of density oscillations by the impurities. This is possible only for specific 
configurations of the impurities, namely, when the pinned density oscillations cancel each other out. 
Other configurations lead to an enhancement of the density oscillations and, simultaneously, 
to a reduction of the current.

The presence of the impurities strongly affects 
the superconducting properties of a nanoring as well. One may not expect
a superconductivity in a system consisting of several lattice sites only,
but the tendency toward the formation of a paired state can be analyzed.
In particular, we have investigated a ring with an attractive on--site
interaction. In order to estimate the strength of the pairing instability 
we have calculated the maximal eigenvalue of the pair--susceptibility--matrix.
This is a gauge invariant quantity that possesses the same space symmetry as the
system and increases with the amplitude of the  pairing potential. We have shown that 
abrupt changes in the persistent current coincide with changes of the pair 
susceptibility. For very large pairing potential all electrons are paired and
the flux dependence of the persistent current is the same as for free carriers 
of charge $2e$. 

In the case of the attractive potential, the presence of impurities
and their configuration are even more important than for a repulsive 
potential. It originates from the fact than for attractive interaction there
is a competition between the CDW and superconductivity
even in the absence of impurities, whereas for $U>0$  
the density--oscillations occur only in the vicinity of impurities. 
We have shown that impurities affect the superconducting properties
indirectly, through an enhancement or a reduction of the CDW order.
There is a single mechanism that determines how impurities affect both the
persistent current for $U>0$ and the pair susceptibility for $U<0$.
Therefore, if a given impurity configuration  leads to an enhancement of the
persistent current for $U>0$, the same configuration leads also to an enhancement 
of the pairing tendency for $U<0$.

The exact diagonalization analysis has been supplemented by the Bogoliubov--de Gennes study
of much larger rings of a finite width. Qualitatively both the approaches give similar 
results concerning the competition between superconductivity and CDW. 
In particular, the flux dependence of the pair
susceptibility in the first case exactly corresponds to that of the pairing amplitude
in the latter case. The role of impurities and their configuration in both the cases 
are the same as well.   
Moreover, the BdG approach allowed us to investigate 
larger systems that exhibit bulk superconductivity and then, reducing their sizes,
to trace how the properties change when entering the nano regime.
The mean--field approximation is generally inappropriate for the low--dimensional systems.
However, a comparison of the results obtained with the help of Lancz{\"o}s and BdG methods
indicates that the mean--field approach gives qualitatively correct results for the
persistent currents in small rings with a weak pairing interaction.

To summarize, we  have demonstrated how imperfectness modifies nanoscopic properties of 
small rings with electronic correlations. This investigations are important in connection
with the recent developments in nanotechnology. In particular, it is possible to fabricate
nanorings with arbitrary configuration of impurities and, in this way, to control the rings' 
properties.

\end{document}